\documentclass[aps,showpacs,preprintnumbers,superscriptaddress,twocolumn,floatfix]{revtex4-1}
\usepackage{amsmath}
\usepackage{eurosym}
\usepackage{amssymb}
\usepackage{graphicx}
\usepackage{color}

\def\be{\begin{equation}}
\def\ee{\end{equation}}
\def\bea{\begin{eqnarray}}
\def\eea{\end{eqnarray}}

\begin{document}

\title{Mass - radius ratio bounds for compact objects in Lorentz-violating dRGT Massive Gravity
theory}
\author{Parinya Kareeso}
\email{Parinya.Kar@student.chula.ac.th}
\affiliation{High Energy Physics Theory Group, Department of Physics, Faculty of Science,
Chulalongkorn University, Phyathai Rd., Bangkok 10330, Thailand}
\author{Piyabut Burikham}
\email{piyabut@gmail.com}
\affiliation{High Energy Physics Theory Group, Department of Physics, Faculty of Science,
Chulalongkorn University, Phyathai Rd., Bangkok 10330, Thailand}
\author{Tiberiu Harko}
\email{t.harko@ucl.ac.uk}
\affiliation{Department of Physics, Babes-Bolyai University, Kogalniceanu Street,
Cluj-Napoca 400084, Romania,}
\affiliation{School of Physics, Sun Yat-Sen University, Guangzhou 510275, People's Republic of China}
\affiliation{Department of Mathematics, University College London, Gower Street, London
WC1E 6BT, United Kingdom}
\date{\today }

\begin{abstract}
We consider the mass-radius bounds for spherically symmetric static compact
objects in the de Rham-Gabadadze-Tolley (dRGT) Massive Gravity theories, free of ghosts. In this type of
gravitational theories the graviton, the quantum of gravity, may have a
small, but non-vanishing mass. We derive the hydrostatic equilibrium and
mass continuity equations in the Lorentz-violating Massive gravity in the presence of a
cosmological constant and for a non-zero graviton mass. The case of the
constant density stars is also investigated by numerically solving the
equilibrium equations. The influence of the graviton mass on the global
parameters (mass and radius) of these stellar configurations is also
considered. The generalized Buchdahl relations, giving the upper and lower
bounds of the mass-radius ratio are obtained, and discussed in detail. As an
application of our results we obtain gravitational redshift bounds for
compact stellar type objects in the Lorentz-violating dRGT Massive Gravity, which may (at least in
principle) be used for observationally testing this theory in an
astrophysical context.

\end{abstract}

\maketitle

\tableofcontents


\section{Introduction}

\label{sect1}

Despite its amazing success in explaining gravitational and cosmological
dynamics on scales ranging from the Solar System to the Hubble radius,
general relativity was confronted from its early stages of existence with a
plethora of alternative gravity theories. An interesting but less
investigated way of explaining gravity was related to field theoretical
models of gravity, in which the gravitational interaction, similarly to the
other interactions of nature, is mediated by a spin two particle, called the
graviton. The early field theoretical approaches to gravity were formulated
in a flat geometry, and the first such model was proposed by Fierz and Pauli
\cite{FP} in 1939. This linear approach to gravity succeeded to give to the
graviton a mass, introduced five degrees of freedom in the model, and
avoided the propagation of the sixth's one. A theory of gravitation using a
massless tensor field was proposed by Thirring in \cite{Th}. In this model
the field equations require a conserved source and admit a gauge-group,
while the equations of motion of particles are gauge invariant only if the
gauge transformation of the field is supplemented by a linear coordinate
transformation.

An important moment in the development of the massive gravity theory was
represented by the paper \cite{vD}, where it was found that there exists a
discrete difference between the zero-mass theories and the very small, but
non-zero mass theories (the vDVZ discontinuity).  In this context it is important to mention that massive gravity is a {\it classical field theory} that does not need to be formulated in terms of the graviton, a particle mediating the gravitational interaction in a way similar to the electromagnetic or nuclear interactions. Based on its {\it transformation properties,  a  classical gravitational field has spin two}, and
 {\it it can have a mass that follows from  its dispersion relation}. These properties are general, and they are valid without the need of introducing a particle representation of the gravitational interaction. In the following we will use, for simplicity,  the term ``graviton" as defined above, and which does not imply an explicit particle interaction picture.
In the case of gravitation,
a comparison of massive and massless theories with experiment, in
particular the perihelion movement of Mercury, did show that the massive
gravity theory must be excluded, and therefore the graviton mass must be
rigorously zero.

A possible way to get around the physical consequences of
the vDVZ discontinuity was proposed in \cite{Vai}, and was based on the idea
that the linearized approximation of the gravitational field breaks down
near massive objects like, for example, the Sun. Therefore an improved
expansion must be used, which, also including the previously ignored
nonlinear effects, leads to a continuous zero mass limit. Static,
spherically symmetric, and asymptotically flat numerical solutions of
massive gravity with a source were obtained in \cite{Bab}, and they led to a
recovery of the Schwarzschild solution of standard general relativity via
the Vainshtein mechanism. The massive gravity theory seemed to face
insurmountable problems after the publication of the paper \cite{B1}, where
it was claimed that no acceptable tensor gravitational theory with
arbitrarily long but finite range could exist. The main points to support
this result are the facts that in the massive version of the full Einstein
theory, there are necessarily six rather than the five tensor degrees of
freedom, the energy has no lower bound, the infinite-range limit does not
exist at all, and lowest-order forces are the same as in the massive
linearized theory, respectively. The Boulware-Deser (BD) ghost instability
raised serious questions about the viability of any massive gravity theory.

However, de Rham, Gabadadze and Tolley (dRGT) \cite{dR1, deRham:2010kj}
succeeded in 2010 to construct the first (and probably unique) nonlinear
fulfillment of the Fierz-Pauli theory that is free of the Boulware-Deser
ghost instability. To achieve this goal the Lagrangian of gravity was
covariantly amended by mass and polynomial interaction terms with arbitrary
coefficients. The consistency of the theory was investigated in the
decoupling limit, up to the fifth order in the nonlinearities. The
ghost-like pathologies in these interactions cancel for special choices of
the polynomial interactions, and it was suggested that this result remains
true to all orders in the decoupling limit. It was also pointed out that the
mixing between the helicity-0 and 2 modes can be at most quartic in the
decoupling limit. The problem of the ghosts in the non-linear massive
gravity was analyzed within the ADM formalism in \cite{Has1,Has2,Has3}, and
it was shown that, in the entire two-parameter family of actions, the
Hamiltonian constraint is maintained at the complete non-linear level. This
result implies the absence of the pathological Boulware-Deser ghost to all
orders. In \cite{dR2} it was shown that there can be no new Lorentz
invariant kinetic interactions free from the Boulware-Deser ghost in four
dimensions in the metric formulation of gravity, beyond the standard
Einstein-Hilbert, up to total derivatives. By performing a general
perturbative analysis in four dimensions, it follows that the only term with
two derivatives that does not introduce a ghost is the Einstein-Hilbert
term. Moreover, this result extends to all orders in perturbations. For
reviews on the theoretical aspects of massive gravity see \cite{rev1}, \cite%
{rev2}, and \cite{rev3}, respectively.

The establishment of a firm theoretical foundation of the massive gravity
theory has opened the possibility of the investigation of its cosmological
and astrophysical applications. The effect of helicity-0 mode which remains
elusive after analysis of cosmological perturbation around an open
Friedmann-Lemaitre-Robertson-Walker universe was investigated in \cite{Chia}%
. The non-linear form of the effective energy-momentum tensor stemming from
the mass term was also derived for the spherically symmetric case. By
solving the spherically symmetric gravitational equations of motion in
vacuum to the linear order, a solution which has an arbitrary time-dependent
parameter was obtained. In general relativity, this parameter corresponds to
the mass of a star. Hence Birkhoff's theorem may no longer hold in the
non-linear massive gravity, and the energy can probably be emitted
superluminously (with infinite speed) on the self-accelerating background by
the helicity-0 mode.

Homogeneous and isotropic cosmological solutions have been presented in \cite%
{cosm1}, which suffer from either Higuchi ghost or a non-linear ghost
instability. By relaxing the symmetry of the background by e.g. breaking
isotropy in the hidden sector, it is possible to accommodate a stable
cosmological solution. Alternatively, extending the theory to allow for new
dynamical degrees of freedom can also remove the conditions that lead to the
instability. The stability of the linear perturbations in the bimetric
theory was examined in \cite{cosm2}. Instabilities were presented for
several classes of models, and simple criteria for the cosmological
stability of massive bigravity were derived. A particular self-accelerating
bigravity model, infinite-branch bigravity, which exhibits both viable
background evolution and stable linear perturbations was also found.  In \cite{cosm2new} it was shown that by taking the Planck mass for the second metric to be small, the instabilities of the bimetric theory describing gravitational interactions in the presence of an extra spin-2 field can be moved back to unobservably early times, when the theory approaches general relativity with an effective cosmological constant determined by the spin-2 interaction scale. The late-time expansion history of the theory becomes extremely close to the standard $\Lambda $CDM model, with a natural value for the cosmological constant. In order for the cosmological perturbations to be stable by Big-Bang nucleosynthesis the Planck mass for the second metric $M_f$ must be smaller than the electroweak scale.

The scalar gravitational radiation from a binary pulsar system in the
simplest model that exhibits the Vainshtein mechanism was computed in \cite%
{dR3}. The gravitational radiation is less suppressed relative to its
general relativity predictions than static fifth forces effects within the
pulsar system. Spherically symmetric solutions of the field equations in the
dRGT massive gravity model have also been extensively investigated. In \cite%
{N1} it was shown that the Schwarzschild-de Sitter and
Reissner-Nordstr\"om-de Sitter black hole metrics appear as exact solutions
in the dRGT model, where the mass term sets the curvature scale. They occur
within a two-parameter family of dGRT mass terms. In the limit of vanishing
graviton mass they go smoothly to the Schwarzschild and Reissner-Nordstr\"om
metrics. Static charged black hole solutions in nonlinear massive gravity
were found in \cite{Cai1}, in the parameter space of two gravitational
potential parameters $(\alpha,\beta)$. In the simplest case with $\alpha
=\beta =0$, the solution exhibits the vDVZ discontinuity but ordinary
General Relativity is recovered deep inside the horizon due to the existence
of electric charge. Spherically symmetric solutions in the bigravity
formulation of massive gravity were obtained in \cite{Com}. The solutions
admit both a Lorentz invariant and a Lorentz breaking asymptotically flat
behaviour and also fall in two branches. In the first branch, all solutions
can be found analytically, and are Schwarzschild-like. In the second branch,
Yukawa-like modifications of the static potential were found.
Spherically-symmetric solutions in Massive Gravity generated by matter
sources with polytropic equation of state were studied in \cite{Bri}, in the
non-perturbative regime where the mass term non-linearities are important. A
detailed study of the spherically symmetric solutions in Lorentz breaking
massive gravity was presented in \cite{Li1}. The stability of the
gravitational field by the analysis of the Komar integral was also
discussed. Static spherically symmetric black hole solutions of dRGT massive
gravity theory in the presence of cosmological constant were obtained in
\cite{Jaf}. The unitary and non-unitary gauges are used to find the
solutions in three, four and five dimensions. Two general classes of
solutions were found, and in the first one the effect of massive potential
appears as the effective cosmological constant. The quasi-stationary profile
of massive charged scalar field in a class of charged black hole in dRGT
massive gravity was investigated in \cite{Bur1}. For asymptotically dRGT
anti de Sitter (AdS) black holes, unstable modes have been found, with their
frequency satisfying the condition of superradiance. The properties of the
black holes in Massive Gravity theory have been investigated in \cite%
{bh1,bh2,bh3,bh4, bh5,bh6,bh7}. 

Relativistic stars in the simplest model of the de Rham-Gabadadze-Tolley massive gravity, which describes the massive graviton without ghost propagating mode were studied in \cite{Neut0}. The modified Tolman-Oppenheimer-Volkoff equation and the constraint equation coming from the potential terms in the gravitational action were derived, and analytical and numerical results for quark and neutron stars were obtained. The deviations were compared with the corresponding results in standard General Relativity and $f(R)$ gravity theory, respectively. The dRGT gravity theory leads to small deviations from the General Relativity in terms of density profiles and mass-radius relation.
The structure of neutron stars in the context of Massive Gravity was studied in \cite{Neut}. The modifications of hydrostatic equilibrium  equation in the presence of massive gravity have been explored in four and higher dimensions.  The consideration of the effects of the Massive Gravity gives specific contributions into the structure of neutron stars. A relation between the mass and radius of neutron stars versus the Planck mass was also obtained.

The study of the stability of compact objects in the general relativistic
framework is of central importance for understanding the behavior of
astrophysical systems such as black holes or neutron stars. A simple but
very powerful stability criterion was obtained by Buchdahl \cite{Bu59,Stra},
and it gives the condition for the stability of a compact object with mass $M
$ and radius $R$ as
\begin{equation}\label{Bstab}
\frac{2GM}{c^2R}\leq \frac{8}{9}.
\end{equation}
 The condition given in Eq. (\ref{Bstab})  is a stability condition in static spherical symmetry against the collapse of massive objects in General Relativity. If the condition is not satisfied, the gravitating object collapses, thus leading to the formation of a black hole. The equality sign (also called the Buchdahl limit) corresponds to the case of constant density stars, and gives the maximum mass-radius ratio for stable massive compact objects. The
Buchdahl upper bound was generalized in \cite{MaDoHa00} to take into account
the effect of the cosmological constant $\Lambda$. Sharp bounds on the
mass-radius ratio were obtained in \cite{An1,An2,An3}. For example, in \cite%
{An3} it was shown that if the energy condition $p+2p_{\perp} \leq \rho $ is
satisfied, where $p\geq 0$ is the radial pressure, and $p_{\perp}$ is the
tangential pressure, then the condition
\begin{equation}
\frac{GM}{c^2R}\leq\frac29-\frac{\Lambda R^2}{3}+\frac29 \sqrt{1+3\Lambda R^2%
},
\end{equation}
must hold. Buchdahl type upper limits for the mass radius ratio have been
obtained for charged particles \cite{MaDoHa01,Boehmer:2007gq,An4}, and for
anisotropic stars \cite{BoHa06}. In \cite{An4} it was shown that for an
object with charge $q$, if the condition $0\leq q^2/r^2+\Lambda r^2\leq 1$
is satisfied, then the inequality
\begin{equation}
\frac{Gm}{c^2r}\leq \frac{2}{9}+\frac{q^2}{3r^2}−\frac{\Lambda
r^2}{3}+\frac{2}{9}\sqrt{1+\frac{3q^2}{r^2}+3\Lambda r^2},
\end{equation}
must hold. Mass-radius ratio bounds were derived for arbitrary dimensional
spheres in \cite{W1}, and for Gauss-Bonnet gravity in \cite{W2}.

A lower bound of the mass-radius ratio in the presence of a cosmological
constant, as well as a cosmological constant related minimum density was
found in \cite{min1}, and further explored in \cite{min2,min3}. These lower
bounds can be formulated as
\begin{equation}
\frac{2GM}{c^{2}R}\geq \frac{1}{6}\Lambda R^2, \rho =\frac{3M}{4\pi R^3}\geq
\rho _{\Lambda}\equiv \frac{\Lambda c^2}{16 \pi G}.
\end{equation}
In the case of a charged particle with total charge $Q$ the lower bound for
the mass-radius ratio is given by \cite{min3}
\begin{equation}
M\geq \frac{3}{4} \frac{Q^2}{Rc^2}+\frac{\Lambda R^3c^2}{12G}.
\end{equation}
By using the minimum mass-cosmological constant relation, as well as
dimensional analysis \cite{Wesson:2003qn}, one can obtain a representation
of the cosmological constant in terms of the fundamental physical constants
as \cite{Not, min2,Beck:2008rd}
\begin{equation}
\Lambda \approx \frac{\hbar ^2G^2m_e^6c^6}{e^{12}},
\end{equation}
where $m_e$ is the electron mass. For a review of the relation between
fundamental physics and the cosmological constant see \cite{L1}.

The mass - radius relations, as well as the possible existence of a minimum
mass have been in different theoretical contexts, and for different physical
models, in \cite{Mass1,Mass2,Mass3,Mass4,Mass5,Mass6}. The generalized
Buchdahl inequalities in arbitrary space-time dimensions in the presence of
a non-zero cosmological constant were obtained in \cite{Mass1}, by
considering both the de Sitter and anti-de Sitter cases. The dependence on
the number of space-time dimensions of the minimum and maximum masses for
stable spherical objects was explicitly obtained. Bounds for the minimum and
maximum mass/radius ratio of a stable, charged, spherically symmetric
compact object in a $D$-dimensional space-times were obtained, in the
presence of dark energy, in \cite{Mass2}. By combining the lower mass bound,
in four space-time dimensions, with minimum length uncertainty relations
(MLUR) motivated by quantum gravity, an alternative bound for the maximum
charge/mass ratio of a stable, gravitating, charged quantum mechanical
object, expressed in terms of fundamental constants, was obtained. This
limit leads to the correct order of magnitude value for the charge/mass
ratio of the electron, as required by the stability conditions. The physical
interpretation of the mass scale $\left(\hbar ^2\sqrt{\Lambda}/G\right)^{1/3}
$ was discussed in \cite{Mass3}. Based on the Generalized Uncertainty
Relation, it was shown that a black hole with age comparable to the age of
the Universe would stop radiating when the mass reaches a new mass scale $%
M^{\prime }_T=c\left(\hbar /G^2\sqrt{\Lambda}\right)^{1/3}$. Upper and lower
bounds on the mass-radius ratio of stable compact objects in extended
gravity theories, in which modifications of the gravitational dynamics are
described by an effective contribution to the matter energy-momentum tensor,
were obtained in \cite{Mass4}. The possibility of a variable coupling
between the matter sector and the gravitational field was considered, and
the obtained results are valid for a large class of generalized gravity
models. As an applications of the obtained formalism compact bosonic
objects, described by scalar-tensor gravitational theories with
self-interacting scalar field potentials, and charged compact objects,
respectively, were considered. By assuming a static, spherically symmetric
geometry, the strong gravity equilibrium properties of compact hadronic
objects were investigated in \cite{Mass5}. The generalized Buchdahl
inequalities for a strong gravity `particles' were derived, and the upper
and lower bounds of the mass/radius ratio of stable, compact, strongly
interacting objects were obtained. The existence of the lower mass bound is
induced by the presence of the effective cosmological constant, which
produces a mass gap, while the upper bound corresponds to a deconfinement
phase transition. Upper and lower limits for the mass-radius ratio of
spin-fluid spheres in Einstein-Cartan theory in the presence of a
cosmological constant were considered in \cite{Mass6}, under the assumption
that matter satisfies a linear barotropic equation of state. In the case of
the spin-generalized strong gravity model for baryons/mesons, show the
existence of quantum spin imposes a lower mass bound for spinning particles,
which almost exactly reproduces the electron mass. The mass-radius relations for neutron stars in $f(R)$ and other modified theories of gravity were investigated in \cite{Cap1,Cap2,Cap3,Cap4}. 

Massive Gravity theories are formulated with the help of a fixed
fiducial metric  $f_{\mu\nu}$,  and the general  properties of the theory
depend very much on the choice of the metric $f_{\mu \nu}$.  Therefore,  each $f$-metric
gives rise to a different massive gravity theory. It is the goal of this work to consider the mass-radius ratio bounds in the
framework of the dRGT  Massive Gravity theory with Lorentz-violating fiducial metric. This represents a generalization of
the previous works on the mass-radius upper and lower bounds to this interesting approach to the gravitational force.
After writing down the gravitational field equations of dRGT Massive Gravity, we
specialize our analysis to the case of the spherically symmetric static
gravitational field. For this particular geometry the hydrostatic
equilibrium equations are obtained, which represent the generalizations of
the standard Tolman-Oppenheimer-Volkoff equation of general relativity, and
of the mass continuity equation, respectively. We investigate through
numerical analysis the solutions of these equations for the simple but
theoretically important case of the constant density stars. The upper and
lower bounds for the mass-radius ratios are obtained, and discussed
systematically for the three possible cases determined by the sign and
numerical value of the parameter $\gamma $ of the model, which is
proportional to the mass square of the graviton. As possible physical
applications of our results we discuss the corrections to the minimum mass
of particles due to the non-zero graviton mass, as well as the modifications
of the surface redshift of the compact gravitational objects.

The present paper is organized as follows. The field equations of the
dRGT Massive Gravity model are introduced in Section~\ref{sect2}, where the
hydrostatic equilibrium equations of compact objects in static spherical
symmetry are derived. The case of the constant density stars is also
investigated. The mass-radius bounds for dense stars are derived in Section~%
\ref{sect3} for arbitrary values of the model parameter $\gamma$. We discuss
and conclude our results in Section~\ref{sect4}. The rescaling of the metric
function is explained in Appendix A.

\section{Field equations, geometry, hydrostatic equilibrium, and constant
density stars in dRGT Massive Gravity}

\label{sect2}

\subsection{The field equations of dRGT Massive Gravity with the Lorentz-violating fiducial metric}

We start with the well-known Einstein-Hilbert gravitational action plus
consistent nonlinear interaction terms interpreted as a graviton mass which
is given by~\cite{deRham:2010kj} 
\begin{equation}
S = \int d^4x \sqrt{-\mathrm{g}}\frac{1}{2\kappa} \bigg[ R+ 2\kappa%
\mathcal{L}_{m} + m^2_{\mathrm{g}}\ \mathcal{U}(\mathrm{g},\phi^a)\bigg] ,
\end{equation}
where $\kappa = 8\pi G/c^4$, $R$ is the scalar curvature, $\mathcal{L}_{m}$
is the matter Lagrangian, and $\mathcal{U}$ is a graviton potential with the
parameter $m_{\mathrm{g}}$ interpreted as graviton mass. The nonlinear
interaction potential, which is constructed to the fourth order in the
four-dimensional spacetime, is given by 
\begin{equation}
\mathcal{U}(\mathrm{g},\phi^a) = \mathcal{U}_{2} + \alpha_{3}\mathcal{U}_{3}
+ \alpha_{4}\mathcal{U}_{4},
\end{equation}
where the coefficients $\alpha_{3}$ and $\alpha_{4}$ are dimensionless free
parameters. The potentials on the second, the third, and the fourth terms
are defined as
\begin{eqnarray}
\hspace{-0.8cm}\mathcal{U}_{2} &\equiv & [\mathcal{K}]^2 - [\mathcal{K}^2], \\
\hspace{-0.8cm}\mathcal{U}_{3} &\equiv & [\mathcal{K}]^3 - 3[\mathcal{K}][\mathcal{K}^2] +
2[\mathcal{K}^3], \\
\hspace{-0.8cm}\mathcal{U}_{4} &\equiv & [\mathcal{K}]^4 - 6[\mathcal{K}]^2[\mathcal{K}^2]
+ 8[\mathcal{K}][\mathcal{K}^3] + 3[\mathcal{K}^2]^2 - 6[\mathcal{K}^4],
\end{eqnarray}
respectively. The building block tensor is defined as 
\begin{equation}\label{Knew}
\mathcal{K}^\mu_\nu = \delta^\mu_\nu - \sqrt{\mathrm{g}^{\mu\sigma}f_{ab}%
\partial_\sigma \phi^a \partial_\nu \phi^b},
\end{equation}
where
\begin{equation}
[\mathcal{K}] = \mathcal{K}^\mu_\mu, [\mathcal{K}^n] = (\mathcal{K}%
^n)^\mu_\mu.
\end{equation}

This choice of interaction eliminates the BD ghost order by order. We follow
the previous works by choosing a simple form of the fiducial metric to be the Lorentz-violating~%
\cite{Berezhiani:2011mt,Ghosh:2015cva}
\begin{equation}\label{14new}
f_{\mu\nu} = \mathrm{diag}(0, 0, \lambda^2, \lambda^2\sin^2\theta),
\end{equation}
where $\lambda$ is a constant, and we choose the unitary gauge $\phi^a =
x^\mu \delta^a_\mu$ for the St\"{u}ckelberg scalars.  In fact, the analysis of \cite{deRham:2010kj} was initially performed for a
flat Minkowski $f$-metric, and the expression of the potential as introduced in \cite{deRham:2010kj}, is valid for such an
$f$ metric.  On the other hand in \cite{Has2} it was shown that the dRGT theory with a generic $f$-metric is also
ghost free, and this result is valid  for the case
of the singular metric (\ref{14new}).  It should be emphasized that this choice of fiducial metric is Lorentz-violating and the resulting massive gravity model is the Lorentz-violating variation of the dRGT model.  The ``1-K'' formulation \cite{Hasnew} is
more convenient to obtain nonlinear solutions,  and it leads more easily to the field equations, 
as well as  to the parameters of the final solution.

In order to simplify the form of the metric, we will reparametrize the
parameters $\alpha_3$ and $\alpha_4$ to two parameters $\alpha$ and $\beta$,
defined by
\begin{equation}
\alpha_3 = \frac{\alpha -1}{3}, \quad\alpha_4 = \frac{\beta}{4} + \frac{%
1-\alpha}{12}.
\end{equation}

After varying the total action $S=S_g+S_m$, where $S_m$ is the matter
action, the modified Einstein field equations in the presence of the
graviton potential are 
\begin{equation}  \label{field}
G_{\mu\nu}-\kappa T_{\mu\nu}+m^2_{\mathrm{g}}X_{\mu\nu}=0,
\end{equation}
where $T_{\mu \nu}$ is the energy-momentum tensor of the matter.

The effective energy-momentum tensor of massive graviton, obtained by
varying the graviton potential term in the action, takes the following form~%
\cite{Berezhiani:2011mt,Ghosh:2015cva}
\begin{eqnarray}
X_{\mu\nu} &=& \mathcal{K}_{\mu\nu} - \mathcal{K }\mathrm{g}_{\mu\nu}  \notag
\\
&&-\alpha\ \Bigg\{\mathcal{K}^2_{\mu\nu} - \mathcal{K}\mathcal{K}_{\mu\nu} +
\frac{[\mathcal{K}]^2 - [\mathcal{K}^2]}{2}\mathrm{g}_{\mu\nu}\Bigg\}  \notag
\\
&&+3\beta\ \Bigg\{\mathcal{K}^3_{\mu\nu} - \mathcal{K}\mathcal{K}^2_{\mu\nu}
+ \frac{1}{2}\mathcal{K}_{\mu\nu}\ \Big\{\lbrack \mathcal{K}]^2 - [\mathcal{K%
}^2]\Big\}  \notag \\
&&-\frac{1}{6}\mathrm{g}_{\mu\nu}\ \Big\{\lbrack \mathcal{K}]^3 - 3[\mathcal{%
K}][\mathcal{K}^2] + 2[\mathcal{K}^3]\Big\}\Bigg\}.
\end{eqnarray}

We will assume that the constraint from Bianchi identities gives separately
the covariant derivatives of $T_{\mu\nu}$ and $X_{\mu\nu}$ equal to zero,
according to the equations
\begin{equation}
\nabla^\mu X_{\mu\nu} = 0,\quad\nabla^\mu T_{\mu\nu} = 0.
\end{equation}

\subsection{The spherically symmetric case}

In four space-time dimensions, we consider a static and spherically
symmetric metric of the following form
\begin{equation}  \label{metric}
ds^2 = -n(r)d(ct)^2 + \frac{dr^2}{f(r)} + r^2d\Omega^2,
\end{equation}
where $d\Omega^2 = d\theta^2+\sin^2\theta d\phi^2$.

We will assume that the energy-momentum tensor of the matter is given by
\begin{equation}
T^{\mu}_{\nu} = (\rho c^2+P)u^{\mu}u_{\nu} + P\mathrm{\delta^{\mu}_{\nu},}
\end{equation}
\textit{i.e.}, by a perfect fluid, characterized by only two thermodynamic
parameters, the matter density $\rho $, and the thermodynamic pressure $P$,
respectively, as well as by its four-velocity $u^{\mu}$, satisfying the normalization condition $u^{\mu}u_{\mu}=-1$. In the following we adopt the comoving reference frame, in which the components of the four velocity are given by $u^{\mu}=\left(-n(r)^{-1/2},0,0,0\right)$.

For the metric given by Eq.~(\ref{metric}), the components of Einstein
tensor become 
\begin{eqnarray}
G^t_t &=& \frac{f^{\prime }}{r}+\frac{f}{r^2}-\frac{1}{r^2}, \\
G^r_r &=& \frac{f(rn^{\prime }+n)}{nr^2}-\frac{1}{r^2}, \\
G^\theta_\theta &=& G^\phi_\phi  \notag \\
&=& f^{\prime }\Big(\frac{n^{\prime }}{4n}+\frac{1}{2r}\Big) + f\Big(\frac{%
n^{\prime \prime }}{2n} + \frac{n^{\prime }}{2nr}-\frac{n^{\prime 2}}{4n^2}%
\Big),
\end{eqnarray}
where a prime denotes the derivative with respect to $r$. The components of
the effective energy-momentum tensor of the massive graviton are given by
\begin{equation}
X^t_t = -\bigg[\frac{\alpha(3r-\lambda)(r-\lambda)}{r^2}+\frac{%
3\beta(r-\lambda)^2}{r^2}+\frac{3r - 2\lambda}{r}\bigg],
\end{equation}
\begin{equation}
X^r_r = -\bigg[\frac{\alpha(3r-\lambda)(r-\lambda)}{r^2}+\frac{%
3\beta(r-\lambda)^2}{r^2}+\frac{3r - 2\lambda}{r}\bigg],
\end{equation}
\begin{equation}
X^\theta_\theta = X^\phi_\phi = \frac{\alpha(2\lambda-3r)}{r}+\frac{%
3\beta(\lambda-r)}{r}+\frac{\lambda-3r}{r}.
\end{equation}

Substitute all components in Eq.~(\ref{field}), the modified Einstein field
equations become

\begin{eqnarray}  \label{Gtt}
\frac{f^{\prime }}{r}+\frac{f}{r^2}-\frac{1}{r^2} &=& m^2_{\mathrm{g}}\bigg[%
\frac{\alpha(3r-\lambda)(r-\lambda)}{r^2}+\frac{3\beta(r-\lambda)^2}{r^2}
\notag \\
&&+\frac{3r - 2\lambda}{r}\bigg] - \frac{8\pi G}{c^2}\rho ,
\end{eqnarray}
\begin{eqnarray}  \label{Grr}
\frac{f(rn^{\prime }+n)}{nr^2}-\frac{1}{r^2} &=& m^2_{\mathrm{g}}\bigg[\frac{%
\alpha(3r-\lambda)(r-\lambda)}{r^2}+\frac{3\beta(r-\lambda)^2}{r^2}  \notag
\\
&&+\frac{3r - 2\lambda}{r}\bigg] + \frac{8\pi G}{c^4}P,
\end{eqnarray}
\begin{eqnarray}
f^{\prime }\Big(\frac{n^{\prime }}{4n}+\frac{1}{2r}\Big) &+& f\Big[\frac{%
n^{\prime \prime }}{2n} + \frac{n^{\prime }}{2nr}-\frac{n^{\prime 2}}{4n^2}%
\Big]  \notag \\
&=& -m^2_{\mathrm{g}}\bigg[\frac{\alpha(2\lambda-3r)}{r}+\frac{3\beta(\lambda-r)}{r%
}  \notag \\
&&+\frac{\lambda-3r}{r}\bigg] + \frac{8\pi G}{c^4}P.
\end{eqnarray}

\subsection{The hydrostatic equilibrium equations}

The functional form of $f$ is obtained from Eq.~(\ref{Gtt}), and can be
expressed as 
\begin{equation}  \label{f-xi}
f(r) = 1 - \frac{2G}{c^2}\frac{M(r)}{r} - \frac{\Lambda}{3}r^2 + \gamma r +
\xi ,
\end{equation}
where
\begin{eqnarray}
\Lambda &=& -3m^2_{\mathrm{g}}(1+\alpha+\beta), \\
\gamma &=& -\lambda m^2_{\mathrm{g}}(1+2\alpha+3\beta), \\
\xi &=& \lambda^2 m^2_{\mathrm{g}}(\alpha+3\beta),  \label{xi}
\end{eqnarray}
and
\begin{equation}
M(r) = 4\pi\int\limits_{0}^{r} \rho(r^{\prime })r^{\prime 2}dr^{\prime },
\end{equation}
respectively, with $M(r)$ representing the total mass inside the radius $r$
of a spherically symmetric object. The graviton mass $m_{\mathrm{g}}$ is
included in the cosmological constant term, namely $\Lambda$, and the extra
terms, $\gamma$ and $\xi$, respectively. The coordinate $r$ can be rescaled
without any loss of generality by setting $\xi =0$ (for details see Appendix
A). Hence, the expression of $f$ can be written as
\begin{equation}  \label{f}
f(r) = 1 - \frac{2G}{c^2}\frac{M(r)}{r} - \frac{\Lambda}{3}r^2 + \gamma r.
\end{equation}
From the continuity equation,$\nabla^\mu T_{\mu\nu} = 0$, it follows that
\begin{equation}  \label{n'/n}
\frac{n^{\prime }}{n} = -\frac{2P^{\prime }}{\rho c^2+P}.
\end{equation}

By substituting Eqs.~(\ref{f}) and~(\ref{n'/n}) in Eq.~(\ref{Grr}), the TOV
equation in the presence of a massive graviton in the dRGT Massive Gravity theory can be obtained as
\begin{equation}  \label{dP/dr}
\frac{dP}{dr} = -\frac{(\rho c^2+P)\Big[\Big(\frac{8\pi G}{c^4}P-\frac{2}{3}%
\Lambda\Big)r^3 + \gamma r^2 +\frac{2G}{c^2}M(r)\Big]}{2r^2\Big[1 - \frac{2G%
}{c^2}\frac{M}{r} - \frac{\Lambda}{3}r^2 + \gamma r\Big]}.
\end{equation}

In order to obtain the structure of stars the hydrostatic equilibrium
equation (\ref{dP/dr}) must be integrated together with the mass continuity
equation
\begin{equation}  \label{dM/dr}
\frac{dM(r)}{dr}=4\pi \rho r^2,
\end{equation}
after the equation of state of the matter, $P=P(\rho)$, was specified. The
boundary conditions that must be imposed at the center and on the surface of
the star are $\rho (0)=\rho _c$, and $P(R)=0$, where $\rho _c$ is the
central density, and $R$ is the radius of the compact object.

The hydrostatic equilibrium and the mass continuity equations can be written
in a dimensionless form with the help of the set of the dimensionless
quantities $\left(\eta, \theta, \Pi\right)$, defined as
\begin{equation}
r=a\eta, M=M^{*}\mu ,\rho =\rho _c\theta, P=\rho _cc^2\Pi,
\end{equation}
where
\begin{equation}
a=\frac{c}{\sqrt{4\pi G\rho _c}}, M^{*}=4\pi \rho _ca^3=\frac{c^3}{\sqrt{%
4\pi G^3\rho _c}}.
\end{equation}

In the new variables the mass continuity and the hydrostatic equilibrium
equations take the form
\begin{equation}  \label{41}
\frac{d\mu}{d\eta}=\theta \eta ^2,
\end{equation}
\begin{equation}  \label{42}
\frac{d\Pi}{d\eta}=-\frac{\left(\theta +\Pi\right)\left[\left(\Pi-\psi%
\right)\eta ^3+\frac{\sigma \eta ^2}{2}+\mu\right]}{\eta ^2\left(1-\frac{2\mu%
}{\eta}-\psi \eta ^2+\sigma \eta\right)},
\end{equation}
where we have denoted
\begin{equation}  \label{42_new}
\psi=\frac{\Lambda}{3}a^2=\frac{\Lambda c^2}{12\pi G\rho _c}, \sigma =\gamma
a=\frac{\gamma c}{\sqrt{4\pi G\rho _c}}.
\end{equation}
In order to close the system of equations (\ref{41}) and (\ref{42}) one must
specify the equation of state of the matter $\Pi=\Pi (\theta)$. The boundary
conditions for the integration of the system are $\theta (0)=1$ and $\Pi
\left(\eta _S\right)=0$, where $\eta _S$ defines the vacuum boundary of the
compact object.

\subsection{Constant density stars in Lorentz-violating dRGT Massive Gravity}

Constant density stars can give in some astrophysical circumstances an
acceptable physical description of realistic astrophysical objects.
Moreover, they are important from theoretical point of view since they allow
some insights into the general properties of the relativistic compact
objects. In the following we will investigate the properties of the constant
density stars in dRGT  Massive Gravity.

\begin{figure}[htb]
\centering
\includegraphics[width=8.5cm]{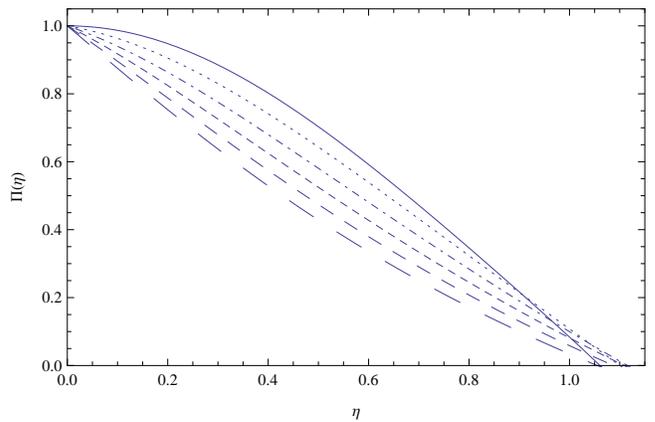}
\caption{Variation of the dimensionless pressure $\Pi$ as a function of the
dimensionless radial coordinate $\eta $ for a constant density star
in dRGT Massive Gravity theory for $\psi =0.06$, and different values of $%
\sigma$: $\sigma =0.25$ (dotted curve), $\sigma =0.50
$ (dashed-dotted curve), $\sigma =0.75$ (short dashed curve), $%
\sigma =1$ (dashed curve), and $\sigma =1.25$ (long dashed
curve). For the sake of comparison we have also presented the standard
general relativistic case, corresponding to $\psi =\sigma =0$
(solid curve). The boundary condition used to integrate the TOV equation are
$\Pi(0)=1$, and $\Pi\left(\eta _s\right)=0$, respectively.}
\label{f1}
\end{figure}

\begin{figure}[htb]
\centering
\includegraphics[width=8.5cm]{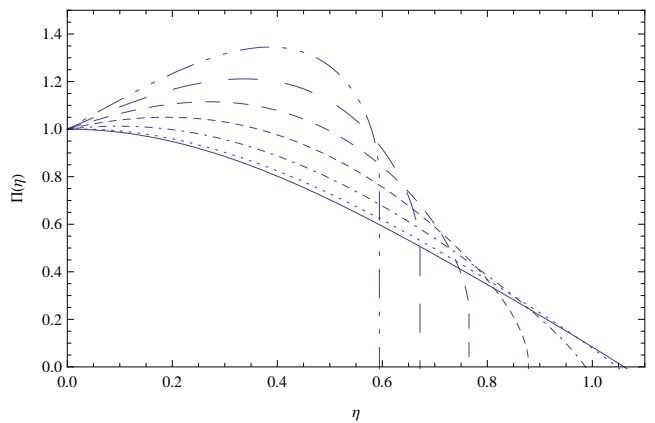}
\caption{Variation of the dimensionless pressure $\Pi$ as a function of the
dimensionless radial coordinate $\eta $ for a constant density star
in dRGT Massive Gravity theory for $\psi =0.06$, and different values of $%
\sigma$: $\sigma =-0.05$ (dotted curve), $\sigma %
=-0.25$ (dashed-dotted curve), $\sigma =-0.50$ (short dashed curve),
$\sigma =-0.75$ (dashed curve), $\sigma =-1$ (long dashed
curve), and $\sigma =-1.25$ (long dashed-double dotted curve). The
standard general relativistic case also corresponds to $\psi =%
\sigma =0$ (solid curve). }
\label{f1.2}
\end{figure}

The requirement of the constant density $\rho =\rho _c=\mathrm{constant},
\forall r\in [0,R]$ fixes the dimensionless density $\theta $ as $\theta =1$
inside the star. Then Eq.~(\ref{41}) can be immediately integrated to give
the dimensionless mass density distribution as
\begin{equation}
\mu (\eta)=\frac{\eta ^3}{3}.
\end{equation}
Substituting this expression of the mass into the hydrostatic equilibrium
equation (\ref{42}) it follows that the pressure $\Pi$ obeys the first order
differential equation given by
\begin{equation}  \label{44}
\frac{d\Pi}{d\eta }=-\frac{\eta \left(1+\Pi\right)\left(\Pi-\psi +\frac{1}{3}%
+\frac{\sigma}{2\eta}\right)}{1-\left(\frac{2}{3}+\psi\right)\eta ^2+\sigma
\eta}.
\end{equation}
Eq.~(\ref{44}) must be integrated with the boundary conditions $\Pi (0)=\Pi
_c$, and $\Pi \left(\eta _S\right)=0$. The variations of the dimensionless
pressure profile inside the constant density star in Massive Gravity theory
are presented for positive and negative numerical values of $\sigma$ in
Figs.~\ref{f1} and \ref{f1.2}, respectively.

As one can see from Fig.~\ref{f1}, the vanishing of the pressure at the
vacuum boundary of the star is dependent on the adopted values of the
positive dimensionless parameter $\sigma$. In the general relativistic case,
corresponding to $\psi =\sigma =0$, the (dimensionless) radius of the star
is given by $\eta _S=1.06$, while $\eta _S=1.10, 1.12, 1.11, 1.09,$ and $1.06
$ for $\sigma =0.25, 0.50, 0.75, 1.00,$ and $1.25$, respectively. This
implies an increase at first, and then a decrease in the radius of the star
with increasing $\sigma $ and for the fixed value $\psi=0.06$.

On the other hand, Fig.~\ref{f1.2} shows the pressure profile inside the sphere for negative $\sigma$. A remarkable feature is the increase of pressure near the central region of the sphere.  It can be shown that this is generic for any negative value of $\sigma$.  For sufficiently small $\eta$, the term $-\sigma/2\eta$ in Eq.~(\ref{44})
becomes dominant and
\begin{equation}
\frac{d\Pi}{d\eta }\simeq-\left(1+\Pi\right)\left(\frac{\sigma}{2}\right),
\end{equation}
Consequently, for sufficiently small $\eta$ and negative~(positive) $\sigma$%
, $\Pi(\eta)$ is always an increasing~(decreasing) function.

At first sight, the increase of the pressure with radius for small $\eta$ seems to imply the existence of an \textit{instability} of this spherical configuration around the center. The reason is we normally need pressure force~($\propto -\vec{\nabla} P$) to exert outward to balance inward gravitational attractive force.  However, for sufficiently small $\eta$, the force of gravity in the massive gravity model with negative $\sigma~(\text{or} ~\gamma)$ actually is {\it always repulsive}, i.e. antigravity.  We can prove this statement by the following.  Generically, the gravity force~(per mass) from the massive gravity metric is given by
\be
-f^{\prime}(r)=\frac{8\pi G r}{c^{2}}(\rho(r)-\frac{\bar{\rho}(r)}{3})+\frac{2\Lambda r}{3}-\gamma.  \label{gforce}
\ee
For sufficiently small $r$, the dominant term is the constant force from massive gravity contribution $\gamma$.  If $\gamma <0$, this force is repulsive, i.e. exerting outwardly from the center of the sphere.  Note also that for sufficiently large $r$, another repulsive ``cosmological constant'' term becomes dominant.

From Eq.~(\ref{gforce}) when $\gamma<0$, the critical radius $r_{c}$ where gravity changes from repulsive in $r<r_{c}$ region to attractive in $r>r_{c}$ region is given by $f^{\prime}(r_{c})=0$~(for constant density profile, there is no $r_{c}$, gravity is always repulsive throughout the object).  On the other hand, Eqn.~(\ref{dP/dr}) tells us that the pressure is an increasing function of radius until
\bea
\frac{8\pi G}{c^{4}}Pr&=&\frac{2\Lambda r}{3}-\gamma-\frac{2GM}{c^{2} r^{2}}, \notag \\
&=&-\frac{8\pi G}{c^{2}}\rho r-f^{\prime}(r),  \label{ngrad}
\eea
then it will start to decrease with respect to $r$.  Therefore, the region of increasing pressure will always be accompanied by antigravity with positive $-f'(r)$ force until $-f'(r)=8\pi G(P+\rho c^{2}) r/c^{4}$ where the pressure starts to decrease with $r$ while gravity is still repulsive.  Beyond this radius, the pressure force becomes repulsive while gravity is still repulsive thus we have instability of the spherical shell.  Interestingly, the static sphere in the negative $\gamma$ scenario is stable with repulsive gravity balancing {\it inward} pressure gradient force!  The radius of the stable compact object in this case is then given by Eqn.~(\ref{ngrad}).  Having high pressure boundary and vacuum outside requires high surface tension for such object to be truly stable under dissipation.

Another interesting possibility of static configuration in negative $\gamma$ scenario is the halo or spherical shell.  This can only occur when $\rho$ is not constant as we can see from Eq.~(\ref{gforce}).  For $r>r_{c}$ in generic profile, Eqn.~(\ref{dP/dr}) guarantees that the pressure gradient force is outward balancing the attractive gravity.  A static halo with inner radius larger than $r_{c}$ is thus stable.

In physical units the radius of the constant density compact objects in dRGT 
Massive Gravity Theory~(for $\sigma\geq 0$) is given by
\begin{equation}
R=10.362\times \left(\frac{\rho _c}{10^{15}\;\mathrm{g/cm^3 }}%
\right)^{-1/2}\times \eta _S\;\mathrm{km}.
\end{equation}
Hence the mass effects associated to the possible existence of the graviton
may change the mass of a neutron star with a central density of the order of
$\rho _c=10^{15}\;\mathrm{g/cm^3}$ from $R\approx 11.0$ km, a value
corresponding to the standard general relativistic case, to $R\approx 11.6$
km, for $\psi =0.06$ and $\sigma =0.50$.

\begin{figure}[htb]
\centering
\includegraphics[width=8.5cm]{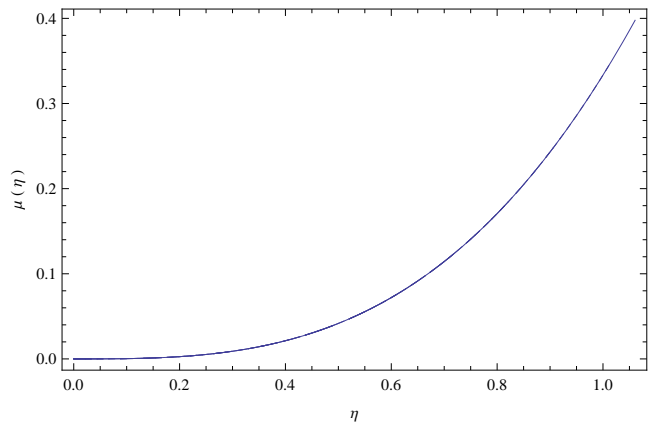}
\caption{Variation of the dimensionless mass $\mu $ as a function of
the dimensionless radial coordinate $\eta $ for a constant density
star in dRGT Massive Gravity theory. }
\label{f2}
\end{figure}

The interior mass profiles of these models are presented in Fig.~\ref{f2}.

For constant density stars all the interior mass profiles follow the same
law, $\mu =\eta ^3/3$. The physical maximum mass $M_S$ of the constant
density star can be obtained as
\begin{equation}
M_S=M^*\mu \left(\eta_S\right)=2.33\times \left(\frac{\rho _c}{10^{15}\;%
\mathrm{g/cm^3 }}\right)^{-1/2}\times \eta _S^3\;M_{\odot}.
\end{equation}
Hence the mass of a constant density star can vary from $M_S=2.78M_{\odot}$
for $\psi =\sigma =0$, corresponding to the general relativistic case, to $%
M_S=3.27M_{\odot}$, corresponding to $\psi =0.06$ and $\sigma =0.50$.

\section{The Buchdahl limits in the Lorentz-violating dRGT Massive Gravity}

\label{sect3}

We introduce now the generalized Buchdahl variables ($x, \omega, \zeta, y$),
defined as follows \newline
\begin{eqnarray}
x=r^2,\omega (r)=\frac{G}{c^2}\frac{M(r)}{r^3},\zeta=n^{1/2}, \\
y^2=f(r)=1-2\omega (r)r^2-\frac{\Lambda}{3}r^2+\gamma r.
\end{eqnarray}

Then Eq.~(\ref{n'/n}) and Eq.~(\ref{dP/dr}) can be rewritten as
\begin{equation}  \label{zeta}
\frac{1}{\zeta}\frac{d\zeta}{dx}=-\frac{1}{\rho c^2+P}\frac{dP}{dx}
\leftrightarrow \frac{d}{dx}(\zeta P)=-\rho c^2\frac{d\zeta}{dx},
\end{equation}
and
\begin{equation}  \label{dP/dx}
\frac{dP}{dx}=-\frac{(\rho c^2+P)}{y^2}\bigg[\frac{2\pi G}{c^4}P-\frac{%
\Lambda}{6}+\frac{\gamma}{4}x^{-1/2}+\frac{\omega}{2}\bigg],
\end{equation}
respectively. By using the above equations, in terms of new variables, we
obtain
\begin{equation}
\frac{d}{dx}\bigg(y\frac{d\zeta}{dx}\bigg)-\frac{1}{2}\frac{\zeta}{y}\frac{%
d\omega}{dx}+\frac{\gamma}{8}\frac{\zeta}{y}x^{-3/2}=0.
\end{equation}

We will introduce a new independent variable $\ell$, obtained by changing
the derivative $2y(d/dx)\rightarrow d/d\ell$, and defined as
\begin{equation}
\ell (r) = \int _{0}^{r} r^{\prime }\bigg[1-\frac{2G}{c^2}\frac{M(r^{\prime
})}{r^{\prime }}-\frac{\Lambda}{3}r^{\prime 2}+\gamma r^{\prime }\bigg]^{-%
\frac{1}{2}}dr^{\prime },
\end{equation}
with $\ell (0)=0$. We define the mean density of the star as
\begin{equation}
\bar{\rho}=\frac{M(r)}{\frac{4}{3}\pi r^3}.
\end{equation}
Both the local density $\rho$ and the mean density $\bar{\rho}$ are required
to be decreasing functions of $r$ inside the spherically symmetric object.
The requirement that the matter density is a decreasing function throughout
the star implies that
\begin{equation}  \label{M}
\frac{d}{dr}\bigg(\frac{M(r)}{r^3}\bigg)< 0,
\end{equation}
leading to $d\omega/dx< 0$.

\subsection{Mass-radius bounds in dRGT Massive gravity for compact objects for $\gamma>0$}


As a first case in the analysis of the mass-radius bounds in dRGT Massive Gravity
we assume the condition that $\gamma>0$. Then for the function $\zeta$ we
obtain immediately the following constraint,
\begin{equation}
\frac{d^2 \zeta}{d\ell^2}<0.
\end{equation}
This condition must hold for all points inside the vacuum boundary of the
compact spherically symmetric object. By using the mean value theorem, we
obtain the inequality
\begin{equation}
\frac{d\zeta}{d\ell}\leq\frac{\zeta (\ell)-\zeta (0)}{\ell -0}\leq\frac{%
\zeta (\ell)}{\ell}.
\end{equation}
Since $\zeta (0)>0$, it follows that
\begin{equation}  \label{bound}
\frac{1}{\zeta}\frac{d\zeta}{d\ell}\leq\frac{1}{\ell}.
\end{equation}
We introduce now the new function $\alpha$(r), defined as
\begin{equation}
\alpha (r)=1-\frac{c^2 r}{2GM}\bigg(-\frac{\Lambda}{3}r^2+\gamma r\bigg).
\end{equation}
This leads to
\begin{equation}
y^2=1-\frac{2G}{c^2}\frac{M(r)}{r}\alpha (r).
\end{equation}
By using the condition ~(\ref{M}), for all $r^{\prime }<r$ we obtain the
inequality,
\begin{equation}
\frac{M(r^{\prime })}{r^{\prime 3}}\geq\frac{M(r)}{r^3}, r^{\prime }<r.
\end{equation}
\newline
Furthermore, we will assume that inside the star the following condition,
\begin{equation}  \label{alpha}
\alpha(r^{\prime })\frac{M(r^{\prime })}{r^{\prime }}\geq\alpha(r)\frac{M(r)%
}{r}\bigg(\frac{r^{\prime }}{r}\bigg)^2,
\end{equation}
also holds. The above inequality follows directly from the requirement,
\begin{equation}
\frac{d}{dr}\Big(\alpha (r)\frac{M(r)}{r^3}\Big)<0,
\end{equation}
leading to
\begin{equation}
\gamma < \frac{8\pi G}{c^2}(\bar{\rho}-\rho)r, r<R,
\end{equation}
which is valid for all $r<R$ where $R$ is the radius of the compact
spherically symmetric object. From Eq.~(\ref{alpha}), it follows that
\begin{equation}
\bigg[1-\frac{2G}{c^2}\frac{M(r^{\prime })\alpha (r^{\prime })}{r^{\prime }}%
\bigg]^{-\frac{1}{2}}\geq\bigg[1-\frac{2G}{c^2}\frac{M(r)\alpha (r)}{r^3}%
r^{\prime 2}\bigg]^{-\frac{1}{2}}.
\end{equation}
Therefore, the right-hand side of inequality~(\ref{bound}) is bounded by
\begin{widetext}
\begin{equation}
 \Bigg\{\int_{0}^{r} r'\bigg[1-\frac{2G}{c^2}\frac{M(r')\alpha(r')}{r'}\bigg]^{-\frac{1}{2}}dr'\Bigg\}^{-1}\leq\frac{2G}{c^2}\frac{M(r)\alpha(r)}{r^3}\Bigg[1-\sqrt{1-\frac{2G}{c^2}\frac{M(r)\alpha(r)}{r}}\Bigg]^{-1}.
\end{equation}
\end{widetext}
The left-hand side of inequality~(\ref{bound}) can be rewritten with the use
of Eqs.~(\ref{zeta}) and~(\ref{dP/dx}), and thus we eventually obtain the
generalized Buchdahl inequality for dRGT Massive Gravity, in the form
\begin{equation}
\bigg(\frac{4\pi G}{c^4}P-\frac{\Lambda}{3}\bigg)r^2+\frac{G}{c^2}\frac{M(r)%
}{r}+\frac{\gamma}{2}r\leq y(1+y).
\end{equation}
This relation is valid for $r\in [0,R]$. The upper bound of the mass-radius
ratio follows by estimating the generalized Buchdahl inequality at the
vacuum boundary of the compact object, where $r=R$, $P(R)=0$, and $M(R)=M$,
respectively. Then it follows that
\begin{widetext}
\bea
\frac{\frac{GM}{c^{2}R}+\frac{\gamma R}{2}-\frac{\Lambda R^{2}}{3}}{\sqrt{1-\frac{2GM}{c^{2}R}-\frac{\Lambda R^{2}}{3}+\gamma R}} \leq \frac{\frac{2GM}{c^{2}R}+\frac{\Lambda R^{2}}{3}-\gamma R}{1-\sqrt{1-\frac{2GM}{c^{2}R}-\frac{\Lambda R^{2}}{3}+\gamma R}}, \label{ineq0}
\eea
\end{widetext}
or in an alternative form
\begin{equation}  \label{ineq}
\frac{3G}{c^2}\frac{M}{R}\leq\sqrt{1-\frac{2G}{c^2}\frac{M}{R}-\frac{\Lambda%
}{3}R^2+\gamma R}+1+\frac{\gamma}{2}R.
\end{equation}
For convenience, the physical variables in the above inequality are
redefined by introducing the dimensionless quantities
\begin{equation}
u:=\frac{G}{c^2}\frac{M}{R}, a:=\frac{\Lambda}{3}R^2, b:=\gamma R.
\end{equation}
Consequently, the inequality~(\ref{ineq}) becomes
\begin{equation}
3u-1-\frac{b}{2}\leq\sqrt{1-2u-a+b}.  \label{ineqdl}
\end{equation}
In order to find the lower and upper bounds on the mass/radius ratio, we
would square the inequality to get rid of the square root. However, this can
be done only when the Left Hand Side  of (\ref{ineqdl}) is bounded from below, i.e. $%
|3u-1-b/2|\leq \sqrt{1-2u-a+b}$. Since the Left Hand Side  is always negative for
sufficiently small $u$~(for large $u$, the Left Hand Side is positive and squaring is
justified naturally), the inequality is trivially satisfied and there is no
lower bound on $u$.

However, there is an additional physical condition to be imposed here. For
the matter sphere to exist physically, we need the matter pressure to
compensate for the pressure from the cosmological constant or the pressure
generated by the massive gravity in this case, i.e.
\begin{equation}
P_{\mathrm{total}}=P+P_{\Lambda}\geq 0
\end{equation}
is required inside the sphere. For negative $P_{\Lambda}$, this condition
implies positive matter pressure $P$. The equation of state of matter thus
demands the matter density $\rho \geq -P_{\Lambda}/wc^{2}>0$ and consequently
\begin{equation}
\bar{\rho}=3M/4\pi R^{3}\gtrsim \Lambda c^{2}/8\pi G  \label{scond}
\end{equation}
inside the sphere. Since $u=4\pi G \bar{\rho}R^{2}/3c^{2}$, the condition $%
|3u-1-b/2|\leq \sqrt{1-2u-a+b}$ is valid when
\begin{equation}
\bar{\rho}\geq \frac{1}{2}\frac{\Lambda c^{2}}{8\pi G},
\end{equation}
\newline
and for small $u$, $a$, and $b$, respectively. This is always true for the
condition (\ref{scond}) above, therefore squaring the inequality is
justified.

By reorganizing the above relation, we obtain
\begin{equation}
9u^2-(4+3b)u+(\frac{b^2}{4}+a)\leq0,
\end{equation}
or, equivalently,
\begin{equation}
(u-u_1)(u-u_2)\leq0,
\end{equation}
where
\begin{equation}
u_1=\frac{4+3\gamma R}{18}\Bigg[1-\sqrt{1-\frac{3\left(3\gamma^2+4\Lambda
\right)R^2}{\left(4+3\gamma R\right)^2}}\Bigg],
\end{equation}
\begin{equation}
u_2=\frac{4+3\gamma R}{18}\Bigg[1+\sqrt{1-\frac{3\left(3\gamma^2+4\Lambda
\right)R^2}{\left(4+3\gamma R\right)^2}}\Bigg].
\end{equation}
Hence, under the condition $\gamma >0$, in the presence of massive graviton,
the mass-radius ratio of compact objects is bounded by
\begin{widetext}
\begin{equation} \label{ineq2}
\frac{4+3\gamma R}{9}\Bigg[1-\sqrt{1-\frac{3\left(3\gamma^2+4\Lambda \right)R^2}{\left(4+3\gamma R\right)^2}}\Bigg]\leq\frac{2GM}{c^2R}\leq\frac{4+3\gamma R}{9}\Bigg[1+\sqrt{1-\frac{3\left(3\gamma^2+4\Lambda \right)R^2}{\left(4+3\gamma R\right)^2}}\Bigg].
\end{equation}
\end{widetext}
The validity of this inequality demands that the value in the square root be
greater than zero, a requirement which leads to the constraint
\begin{equation}
\Lambda<\frac{4+6\gamma R}{3R^2}.  \label{boundcond}
\end{equation}
Nontrivial (positive) lower bounds do exist only when the fraction in the
square root is greater than zero, which gives another constraint for the
negative $\Lambda$ case,
\begin{equation}
\gamma>\sqrt{-\frac{4\Lambda}{3}}, \Lambda<0,
\end{equation}
whilst it is trivially satisfied for $\Lambda >0$ as long as (\ref{boundcond}%
) is valid.

\subsection{Mass-radius ratios in the presence of a cosmological constant
only: the case $\gamma=0$}

For the case $\gamma=0$, Eq.~(\ref{ineq2}) leads to the condition for the
existence of a lower and an upper mass-radius bound, which is given by
\begin{equation}  \label{lambdaineq}
\frac{4}{9}\Bigg[1-\sqrt{1-\frac{3}{4}\Lambda R^2}\Bigg]\leq\frac{2GM}{c^2R}%
\leq\frac{4}{9}\Bigg[1+\sqrt{1-\frac{3}{4}\Lambda R^2}\Bigg].
\end{equation}
This relation implies the existence of a minimum mass/radius ratio for a
matter particle, which is induced by the presence of a cosmological
constant, as shown first in \cite{min1}. The existence of $\Lambda$ also
determines modifications of the Buchdahl upper limit of general relativity
\cite{MaDoHa00}. The generalized Buchdahl inequality (\ref{lambdaineq})
gives a nontrivial solution only when the condition
\begin{equation}
0<\Lambda<\left(\frac{4}{3}\right)R^{-2},
\end{equation}
corresponding to a Schwarzschild-de Sitter type geometry, is satisfied.

\subsection{Mass-radius bounds in dRGT Massive Gravity for dense stars for $\gamma<0$}

For the case $\gamma < 0$, we can write the generalized Buchdahl equation
for spherically symmetric objects in the following form
\begin{equation}  \label{B2}
y(y\zeta^{\prime })^{\prime }=\frac{1}{2}\omega^{\prime }\zeta+\frac{|\gamma|%
}{8}\frac{\zeta}{x^{3/2}}.
\end{equation}
Subsequently, we introduce four new variables $\Gamma, \psi, \eta$ and $z$,
defined as
\begin{equation}
\Gamma(r)\equiv\frac{|\gamma|}{8}\frac{\zeta}{r^2},
\end{equation}
\begin{equation}
\psi=\zeta-\eta,
\end{equation}
where
\begin{equation}
\eta=4\int_{0}^{r} \Bigg(\int_{0}^{r_{1}} \frac{\Gamma(r_{2})}{\sqrt{1-\frac{%
\Theta(r_{2})}{r_{2}}}}dr_{2}\Bigg)\frac{r_{1}}{\sqrt{1-\frac{\Theta(r_{1})}{%
r_{1}}}} dr_{1},  \label{var_new_1}
\end{equation}
while the last variable $z$ is given by
\begin{equation}
dz=\frac{1}{y(x)}dx \rightarrow z(r)=\int_{0}^{r} \frac{2r^{\prime }}{\sqrt{%
1-\frac{\Theta(r^{\prime })}{r^{\prime }}}}dr^{\prime }.  \label{var_new_2}
\end{equation}
The function $\Theta(r)$ is obviously defined by
\begin{equation}
y^2=1-\frac{\Theta(r)}{r},
\end{equation}
where
\begin{equation}
\Theta(r)=\frac{2GM(r)}{c^2}+\frac{\Lambda}{3}r^3+|\gamma|r^2.
\end{equation}
\newline
In terms of the new variables defined above, the Buchdahl equation Eq.~(\ref%
{B2}) can be written as
\begin{equation}
\frac{d^2 \psi(z)}{dz^2}=\frac{1}{2}\omega^{\prime }(x)\zeta(x).
\end{equation}
We assume first the condition that, for $r^{\prime }<r$,
\begin{equation}  \label{Theta}
\frac{\Theta(r^{\prime })}{r^{\prime }}\geq\frac{\Theta(r)}{r}\bigg(\frac{%
r^{\prime }}{r}\bigg)^2,r^{\prime }<r,
\end{equation}
and use the assumption that the density inside the object does not increase
with $r$ in the above relation. Finally, the above assumptions lead to the
condition
\begin{equation}
|\gamma| > -\frac{8\pi G}{c^2}(\bar{\rho}-\rho)r,
\end{equation}
which is valid for all $r\leq R$ as long as the matter density is a
decreasing function of the radial coordinate $r$. Next, as a second
condition we assume that for $r^{\prime }<r$,
\begin{equation}  \label{Gamma}
\Gamma(r^{\prime })\geq\Gamma(r),
\end{equation}
that is, $\Gamma(r)$ is a decreasing function of $r$. This condition leads
to a constraint on $|\gamma|$ given by
\begin{equation}
|\gamma| < \frac{4}{3r}- \frac{40\pi G}{9c^2}\bar{\rho}r+\frac{2}{9}\Lambda
r-\frac{8\pi G}{3c^{4}}Pr.
\end{equation}
This relation is trivially valid for $r\to 0$, and at the surface $r=R$ it
gives a constraint
\begin{equation}
|\gamma| < \frac{4}{3R}- \frac{10G}{3c^2}\frac{M}{R^{2}}+\frac{2}{9}\Lambda
R.
\end{equation}
Alternatively, it can be written with the help of the dimensionless
parameters already defined in Eq.~(\ref{42_new}) as
\begin{equation}
\sigma>-\frac{4}{3\eta}+\frac{10}{9}\eta-\frac{2}{3}\psi\eta.
\end{equation}
For example, when $\psi$ is equal to 0.06, all cases with $\sigma = -0.05,
-0.25, -0.50, -0.75, -1$, and $-1.25$ satisfy this condition at the surface.

From the condition $\omega^{\prime }(x)<0$, we obtain the inequality
\begin{equation}
\frac{d^2}{dz^2}\psi(z)<0,
\end{equation}
which holds for all $r$ in the range $0\leq r\leq R$. Again, by using the
mean value theorem, we find
\begin{equation}
\frac{d\psi}{dz}\leq\frac{\psi(z)-\psi(0)}{z}\leq\frac{\psi(z)}{z}.
\end{equation}
Since $\psi(0)=\zeta(0)-\eta(0)=\zeta(0)>0$, it follows that
\begin{equation}
\frac{d\psi}{dz}\leq\frac{\psi(z)}{z}\rightarrow\frac{d\zeta}{dz}-\frac{d\eta%
}{dz}\leq\frac{\zeta-\eta}{z}.
\end{equation}
After substituting the new variables (\ref{var_new_1}) and (\ref{var_new_2})
in the above relation, we obtain
\begin{widetext}
\begin{eqnarray}
&&\frac{1}{2r}\sqrt{1-\frac{\Theta(r)}{r}}\frac{d\zeta}{dr}-2\int_{0}^{r}\frac{\Gamma(r')}{\sqrt{1-\frac{\Theta(r')}{r'}}}dr'\leq \frac{1}{2\int_{0}^{r}\frac{r'}{\sqrt{1-\frac{\Theta(r')}{r'}}}dr'}\Bigg[\zeta-4\int_{0}^{r}\frac{r_{1}}{\sqrt{1-\frac{\Theta(r_{1})}{r_{1}}}}\Bigg(\int_{0}^{r_{1}}\frac{\Gamma(r_{2})}{\sqrt{1-\frac{\Theta(r_{2})}{r_{2}}}}dr_{2}\Bigg)dr_{1}\Bigg].  \notag \\  \label{76}
\end{eqnarray}
\end{widetext}

The denominator of the right-hand side of Eq.~(\ref{76}) is bounded from
above as a result of using the condition ~(\ref{Theta}). Hence we have
\begin{equation}  \label{77}
\Bigg(\int_{0}^{r} \frac{r^{\prime }}{\sqrt{1-\frac{\Theta(r^{\prime })}{%
r^{\prime }}}}dr^{\prime }\Bigg)^{-1} \leq \frac{\Theta(r)}{r^3}\Bigg(1-%
\sqrt{1-\frac{\Theta(r)}{r}}\Bigg)^{-1}.
\end{equation}

As for the term related to $\Gamma$, it is also bounded as a consequence of
the conditions ~(\ref{Theta}) and ~(\ref{Gamma}), such that
\begin{eqnarray}  \label{79}
\int_{0}^{r} \frac{\Gamma(r^{\prime })}{\sqrt{1-\frac{\Theta(r^{\prime })}{%
r^{\prime }}}}dr^{\prime }&\geq& \Gamma(r)\int_{0}^{r} \Bigg(1-\frac{%
\Theta(r)}{r}\bigg(\frac{r^{\prime }}{r}\bigg)^2\Bigg)^{-\frac{1}{2}%
}dr^{\prime }  \notag \\
&=&\Gamma(r)\bigg(\frac{\Theta(r)}{r^3}\bigg)^{-\frac{1}{2}}\arcsin\Bigg(%
\sqrt{\frac{\Theta(r)}{r}}\Bigg).  \notag \\
\end{eqnarray}
\begin{widetext}
Hence the term in the numerator on the right-handed side of Eq.~(\ref{76}) has a lower bound, which can be obtained as
\begin{eqnarray} \label{80}
&&\int_{0}^{r} \frac{r_{1}}{\sqrt{1-\frac{\Theta(r_{1})}{r_{1}}}}\Bigg(\int_{0}^{r_{1}}\frac{\Gamma(r_{2})}{\sqrt{1-\frac{\Theta(r_{2})}{r_{2}}}}dr_{2}\Bigg)dr_{1} \geq \int_{0}^{r} r_{1}\bigg(1-\frac{\Theta(r_{1})}{r_{1}}\bigg)^{-\frac{1}{2}}\left(\frac{\Gamma(r)}{\Big(\frac{\Theta(r)}{r^3}\Big)^{\frac{1}{2}}}\arcsin\left(\sqrt{\frac{\Theta(r)}{r}}\right)\right)dr_{1}
\notag \\
&&\geq \int_{0}^{r} r_{1}\bigg(1-\frac{\Theta(r)}{r^3}r_{1}^2\bigg)^{-\frac{1}{2}}\left(\frac{\Gamma(r)}{\Big(\frac{\Theta(r)}{r^3}\Big)^{\frac{1}{2}}}\arcsin\left(\sqrt{\frac{\Theta(r)}{r}}\right)\right)dr_{1}
\notag \\
&&= \left(\frac{1-\sqrt{1-\frac{\Theta(r)}{r}}}{\frac{\Theta(r)}{r^3}}\right)\frac{\Gamma(r)}{\Big(\frac{\Theta(r)}{r^3}\Big)^{\frac{1}{2}}}\arcsin\left(\sqrt{\frac{\Theta(r)}{r}}\right)=\frac{\Gamma(r)}{\Big(\frac{\Theta(r)}{r^3}\Big)^{\frac{3}{2}}}\left[\arcsin\left(\sqrt{\frac{\Theta(r)}{r}}\right)-\sqrt{1-\frac{\Theta(r)}{r}}\arcsin\left(\sqrt{\frac{\Theta(r)}{r}}\right)\right]
\notag \\
&&\geq \frac{\Gamma(r)}{\Big(\frac{\Theta(r)}{r^3}\Big)^{\frac{3}{2}}}\left[\sqrt{\frac{\Theta(r)}{r}}-\sqrt{1-\frac{\Theta(r)}{r}}\arcsin\left(\sqrt{\frac{\Theta(r)}{r}}\right)\right],
\end{eqnarray}
\end{widetext}
where we have used the identity $\arcsin x \geq x$. Subsequently, we insert
the inequalities ~(\ref{77}), ~(\ref{79}) and ~(\ref{80}) into Eq.~(\ref{76}%
) and use the relation $y^2=1-\Theta(r)/r$. Afterwards, we obtain
\begin{widetext}
\begin{equation}
\frac{y}{r}\frac{d\zeta}{dr}\leq \frac{1+y}{r^2}\left[\zeta(r)-4\Gamma(r)r^3\left(\frac{1}{1-y^2}-\frac{y\arcsin(\sqrt{1-y^2})}{(1-y^2)^{3/2}}\right)\right]+4\Gamma(r)\frac{r}{\sqrt{1-y^2}}\arcsin(\sqrt{1-y^2}).
\end{equation}
Since for ordinary matter the energy condition $\rho c^2+P\geq0$ always holds, it allows us to replace $1/\zeta$ with $1/y$, such that
\begin{eqnarray}
\left(\frac{4\pi G}{c^4}P-\frac{\Lambda}{3}\right)r^2+\frac{G}{c^2}\frac{M(r)}{r}-\frac{|\gamma|}{2}r &\leq& y(1+y)+\frac{4\Gamma(r)r^3}{1-y}\left(\frac{y\arcsin(\sqrt{1-y^2})}{\sqrt{1-y^2}}-1\right)+4\Gamma(r)r^3\frac{\arcsin(\sqrt{1-y^2})}{\sqrt{1-y^2}}
\notag \\
&\leq& y(1+y)+4r^3\frac{\Gamma(r)}{y},
\end{eqnarray}
\end{widetext}
where we have also used the relation
\begin{equation}
\arcsin(\sqrt{1-y^2}) \leq \frac{\sqrt{1-y^2}}{y}.
\end{equation}
Hence we have obtained the Buchdahl inequality for the mass - radius ratio
of a compact object in dRGT Massive Gravity theory for the case $\gamma<0$. This
inequality is valid for all values of the radial coordinate inside the star,
$r\in [0,R]$. The upper and lower bounds on the mass-radius ratio are
determined by considering the Buchdahl inequality at the surface of the
object, where $r=R$, $P(R)=0$ and $M(R)=M$, respectively. Then it follows
that
\begin{equation}  \label{ineq3}
\frac{3G}{c^2}\frac{M}{R}\leq\sqrt{1-\frac{2G}{c^2}\frac{M}{R}-\frac{\Lambda%
}{3}R^2 - |\gamma| R}+1.
\end{equation}
For convenience, the variables in the above inequality are redefined as
follows
\begin{equation}
u=\frac{G}{c^2}\frac{M}{R}, a=\frac{\Lambda}{3}R^2, b=|\gamma| R.
\end{equation}
Consequently, the inequality~(\ref{ineq3}) becomes
\begin{equation}
3u\leq\sqrt{1-2u-a-b}+1.
\end{equation}
After squaring and simplifying the above relation, we obtain
\begin{equation}
u^2-\frac{4u}{9}+\frac{(a+b)}{9}\leq0,
\end{equation}
or, equivalently,
\begin{equation}
(u-u_{-})(u-u_{+})\leq0,
\end{equation}
where
\begin{equation}
u_{\pm}=\frac{2}{9}\left[1\pm\sqrt{1-\frac{3\left(\Lambda R+3|\gamma|\right)R%
}{4}}\right].
\end{equation}

Hence we have obtained the following lower and upper bounds for the
mass-radius ratio of compact objects in dRGT Massive Gravity,
\begin{eqnarray}  \label{111}
&&\frac{4}{9}\left[1-\sqrt{1-\frac{3\left(\Lambda R+3|\gamma|\right)R}{4}}%
\right]\leq \frac{2GM}{c^2R}\leq  \notag \\
&& \frac{4}{9}\left[1+\sqrt{1-\frac{3\left(\Lambda R+3|\gamma|\right)R}{4}}%
\right].
\end{eqnarray}
The inequality demands the value in the square root greater than zero which
leads to a constraint
\begin{equation}
|\gamma| < \frac{4}{9R}-\frac{\Lambda}{3}R.
\end{equation}
A nontrivial (positive) lower bound in this case exists only when the
fraction in the square root is greater than zero giving another constraint
\begin{equation}
|\gamma| > -\frac{\Lambda}{3}R.
\end{equation}

\section{Discussions and final remarks}

\label{sect4}

Massive Gravity is an interesting theory of gravitation, inspired by the
quantum field theoretical approach to gravity, and which assumes a non-zero
mass of the quanta intermediating the gravitational interaction, the
graviton. Despite the initial many complicated theoretical problems raised
by this approach, a consistent formulation proposed in \cite{dR1} and \cite%
{deRham:2010kj} seems to offer the possibility of an alternative to standard
general relativity, which allows us to go beyond the theoretical limits
imposed by Einstein's theory. The dRGT model of Massive Gravity is ghost-free,
and, at least at the classical level, it has strictly five (or seven in the
bimetric case) gravitational degrees of freedom \cite{Has1,Has2,Has3}.
However, when applied to cosmology, it turns out that the theory with
Minkowski fiducial metric does not have flat and closed
Friedmann-Lemaitre-Robertson-Walker solutions \cite{cosm}. In the context of
the very early Universe cosmology, that is, during inflation, the
propagation of the gravitational waves would also be affected by the
non-trivial mass of the graviton \cite{cosm1a,cosm2a,cosm3,cosm4}.

In the present paper we have investigated in the framework of Lorentz-violating dRGT Massive Gravity theory an important property of compact
general relativistic objects, namely, their mass - radius ratio bounds,
which are important indicators of their stability properties. These bounds
can be obtained from the generalized Buchdahl inequality, from which the
existence of a minimum value of this ratio, as well as an upper stability
limit do follow. In order to obtain the mass-radius ratio bounds we have
adopted a specific form for the $g_{11}$ component of the metric tensor, in
which the corrections to the standard Schwarzschild-de Sitter geometry are
represented by a correction term of the form $\gamma r$, where the
coefficient $\gamma $, proportional to the graviton mass square, gives the new
contribution coming from the ghost-free massive gravity. After adopting the
functional form of the metric, we have obtained the basic equations
describing the hydrostatic equilibrium properties of high density stars. As
compared to the standard general relativistic case, a new term of the form $%
\gamma r^2$, depending on the mass of the graviton, does appear in the TOV
equation. We have first investigated the role this term may play in the
description of stellar properties for the case of constant density stars. In
some astrophysical situations the assumption of constant density may give a
good description of the global parameters of high density objects. As
opposed to the standard general relativistic case, in dRGT Massive Gravity theory
there is no exact solution of the gravitational field equations, and hence a
numerical investigation is required. The pressure distribution inside the
star, and consequently its radius, shows a significant theoretical
dependence on the numerical values of the dimensionless parameter $\sigma$,
constructed from $\gamma $, the density of the star, and the fundamental
constants of physics.

We also explored the stability of a static sphere in dRGT massive gravity model.
Interestingly, the linear term $\gamma r$ in the metric has a crucial role
in the stability condition. When $\gamma$ is negative, the massive-gravity TOV equation demonstrates the universal gravitational stability of a static sphere between {\it repulsive} gravity and inward pressure gradient force in contrast to the conventional gravitational stability of compact object.  Such object, however, requires high surface tension to maintain the high pressure boundary condition.  Interestingly enough,
stable static hollow spheres or halo configurations are also possible for the $%
\gamma <0$ case as long as the inner radius is larger than the turnover radius $r_{c}$ of gravity.

We have obtained, and investigated in detail the Buchdahl inequality for
both a positive and negative $\gamma $ (the case $\gamma =0$ reduces the
model to the standard general relativistic case). In the case $\gamma >0$,
the Buchdahl inequality implies the existence of an absolute minimum
particle mass, which is given by
\begin{equation}
\frac{2GM}{c^2R}\geq \frac{\gamma ^2}{8}\frac{\left(1+4\Lambda /3 \gamma
^2\right)R^2}{\left(1+3\gamma R/4\right)}, \gamma >0.
\end{equation}
Alternatively, this relation can be formulated in terms of an absolute
minimum density $\rho _{min}$, so that the density $\rho =3M/4\pi R^3$ of
any matter configuration must satisfy the constraint
\begin{equation}
\rho \geq \rho _{min}\equiv \frac{3c^{2}\gamma ^2}{64\pi G}\frac{%
\left(1+4\Lambda /3 \gamma ^2\right)}{\left(1+3\gamma R/4\right)}, \gamma >0.
\end{equation}
However, the minimum density is radius-dependent, and the above inequality
can also be interpreted as a matter density-radius relation. It is important
to mention that a lower limit for the mass does exist in massive gravity
even in the absence of the cosmological constant, when $\Lambda =0$. In this
case we have
\begin{equation}
\frac{2GM}{c^2R}\geq \frac{\gamma ^2}{8}\frac{R^2}{\left(1+3\gamma R/4\right)%
}, \gamma >0,
\end{equation}
and
\begin{equation}
\rho \geq \rho _{min}\equiv\frac{3c^{2}\gamma ^2}{64\pi G}\frac{1}{%
\left(1+3\gamma R/4\right)}, \gamma >0,
\end{equation}
respectively. Therefore there is a straightforward relation between the
minimum mass an elementary particle can have, and the mass of the graviton.
If a quantum of gravity does exist, its existence would impose a strong
limit on the minimum mass a particle can have. From a physical point of view
one can assume that it is the graviton mass that determines the
gravitational mass of the elementary particles, and mediates their
gravitational interactions.

A very different minimum mass expression is obtained in the case $\gamma <0$%
. From Eq.~(\ref{111}) we immediately obtain
\begin{equation}
\frac{2GM}{c^2R}\geq \frac{|\gamma |\left(1+\Lambda R/3|\gamma |\right)R}{2}%
, \gamma <0.
\end{equation}
As for the particle mass density, it satisfies a lower bound given by
\begin{equation}
\rho \geq \rho _{min}\equiv \frac{3c^{2}|\gamma | }{16\pi G}\frac{%
\left(1+\Lambda R/3|\gamma| \right)}{R}, \gamma <0.
\end{equation}
Similarly to the $\gamma >0$ case, a minimum mass does exist even in the
absence of the cosmological constant, $\Lambda =0$, and it is given by
\begin{equation}
\frac{2GM}{c^2R}\geq \frac{|\gamma |R}{2}, \gamma <0.
\end{equation}
A similar relation is obtained if the condition $\Lambda R/3|\gamma |<<1$ is
satisfied for all $R$. As for the minimum matter density, it is given by a
relation of the form
\begin{equation}
\rho _{min}\equiv \frac{3c^{2}|\gamma | }{16\pi G R}, \gamma <0.
\end{equation}

As for the upper bounds of the mass-radius ratios of the compact stars in
massive gravity, they are given by
\begin{equation}
\frac{2GM}{c^2R}\leq \frac{4}{9}\left(1+\frac{3\gamma R}{4}\right)\left[2-%
\frac{9\gamma ^2\left(1+4\Lambda /3\gamma ^2\right)R^2}{32\left(1+3\gamma
R/4\right)^2}\right], \gamma >0,
\end{equation}
and
\begin{equation}
\frac{2GM}{c^2R}\leq \frac{4}{9}\left[2-\frac{3\left(\Lambda
R+3|\gamma|\right)R}{8}\right], \gamma <0.
\end{equation}
In both cases in the limit $\gamma =0$, $\Lambda =0$, the corresponding
expressions reduce to the standard Buchdahl limit $2GM/c^2R\leq 8/9$.

The existence of upper/lower bounds of the mass-radius ratio for compact
objects also leads to the existence of some limiting values for other
physical and geometrical quantities of observational interest. One such
important quantity is the surface red shift $z$ of the high density star,
which can be defined in the massive gravity effects corrected
Schwarzschild-de Sitter geometry as
\begin{equation}
z\equiv \frac{1}{\sqrt{f(r)}}-1=\frac{1}{\sqrt{ 1 - 2GM(r)/c^2r - \Lambda
r^2/3 + \gamma r}}-1.
\end{equation}
We consider first the case $\gamma >0$. Then, from Eq.~(\ref{ineq0}),
written as
\begin{equation}
\frac{1}{y}\left[-\frac{\Lambda}{3}R^2+\frac{GM}{c^2R}+\frac{\gamma R}{2}%
\right]\leq \frac{1}{1-y}\left[\frac{2GM}{c^2R} + \frac{\Lambda}{3}R^2 -
\gamma R\right],
\end{equation}
we obtain
\begin{equation}
z\leq \frac{2GM/c^2R-\gamma R+\Lambda R^2/3}{GM/c^2R+\gamma R/2-\Lambda R^2/3%
}.
\end{equation}

In the case $\gamma \equiv 0$ and $\Lambda \equiv 0$, we reobtain the
standard general relativistic gravitational redshift restriction $z\leq 2$.
Alternatively, the redshift bound can be reformulated as
\begin{equation}
z\leq \frac{2\left[ 1-\left( c^{2}/8\pi G\bar{\rho}\right) \left( 3\gamma
/R-\Lambda \right) \right] }{1+\left( c^{2}/8\pi G\bar{\rho}\right) \left(
3\gamma /R-2\Lambda \right) }.
\end{equation}
Hence, at least in principle, observations of the gravitational redshift from compact high density astrophysical objects may offer the possibility of discriminating between Massive Gravity and other modified theories of gravity.

To conclude, in the present paper we have investigated some of the implications of the dRGT Massive Gravity theory with Lorentz-violating fiducial metric, which are relevant at both microscopic and macroscopic scale.   The results obtained in the present analysis may provide some insights for the possible experimental/observational testing of this particular class of Massive Gravity theory at both elementary particle and astrophysical levels, as well as on its theoretical foundations.

\section*{Acknowledgments}

We would like to thank the anonymous reviewer for comments and suggestions that helped us to improve our manuscript. P. K. is supported in part by Graduate School Thesis Grant, Chulalongkorn University. P.B. is supported in part by the Thailand Research Fund~(TRF), Office of Higher Education Commission (OHEC) and Chulalongkorn University under grant RSA6180002. T. H. would like to thank the Yat-Sen School of the Sun Yat-Sen University, Guangzhou, P. R. China, for the kind hospitality offered during the preparation of this work.

\appendix

\section*{Appendix A: Rescaling of the metric and the value of $\xi$}

\label{app}

The vacuum spherically symmetric metric $f$ in massive gravity is given by
Eq.~(\ref{f-xi}), and has the form
\begin{equation}
f(r) = 1 - \frac{2G}{c^2}\frac{M}{r} - \frac{\Lambda}{3}r^2 + \gamma r + \xi
.
\end{equation}
We can rescale the coordinate $r$ by setting
\begin{equation}
r^{\prime }=r/\sqrt{1+\xi},
\end{equation}
leading to
\begin{eqnarray}
ds^2&=&-n(r^{\prime})d(ct)^{2}+\frac{dr^{\prime 2}}{1-\frac{2G}{c^2}\frac{%
M(r^{\prime })}{r^{\prime }}-\frac{\Lambda}{3}r^{\prime 2}+\frac{\gamma}{%
\sqrt{1+\xi}}r^{\prime }}  \notag \\
&&+r^{\prime 2}(1+\xi)d\Omega^2.
\end{eqnarray}

By considering a small spherical surface, its area is given by $%
4\pi(1+\xi)r^{\prime 2}$. However, the surface of the sphere with radius
ranging from Solar System scales up to extragalactic or cosmological scale
is very close to $4\pi r^2$, i.e., the Universe is spatially flat.
Therefore, we set the value of $\xi$ to be zero. Accordingly, the metric $f$
becomes
\begin{equation}
f(r) = 1 - \frac{2G}{c^2}\frac{M}{r} - \frac{\Lambda}{3}r^2 + \gamma r .
\end{equation}
Nevertheless, with the use of Eq.~(\ref{xi}) it follows that the value $\xi=0
$ leads to the condition $\alpha=-3\beta$. This condition affects both $%
\gamma$ and $\Lambda$, but $\gamma$ also depends on $\lambda$. Consequently,
$\gamma$ and $\Lambda$ are the two remaining independent parameters in the
metric.


\begin{thebibliography}{99}
\bibitem{FP} M. Fierz and W. Pauli, Proc. Roy. Soc. Lond. \textbf{A 173},
211 (1939).

\bibitem{Th} W. E. Thirring, Annals of Physics \textbf{16}, 96 (1961).

\bibitem{vD} H. van Dam and M. Veltman, Nuclear Physics \textbf{B 22}, 397
(1970).

\bibitem{Vai} A. Vainshtein, Phys. Lett. \textbf{B 39}, 393 (1972).

\bibitem{Bab} E. Babichev, C. Deffayet, and R. Ziour, Phys. Rev. Lett.
\textbf{103}, 201102 (2009).

\bibitem{B1} D. Boulware and S. Deser, Phys. Rev. \textbf{D 6}, 3368 (1972).

\bibitem{dR1} C. de Rham and G. Gabadadze, Phys. Rev. \textbf{D 82}, 044020
(2010).

\bibitem{deRham:2010kj}  C.~de Rham, G.~Gabadadze and A.~J.~Tolley,
Phys.\ Rev.\ Lett.\ \textbf{106}, 231101 (2011).

\bibitem{Has1} S. Hassan and R. A. Rosen, Phys. Rev. Lett. \textbf{108},
041101 (2012).

\bibitem{Has2} S. Hassan, R. A. Rosen, and A. Schmidt-May, JHEP \textbf{1202}%
, 026 (2012).

\bibitem{Has3} S. Hassan and R. A. Rosen, JHEP \textbf{1202}, 126 (2012).

\bibitem{dR2} C. de Rham, A. Matas, and A. J. Tolley, Classical and Quantum
Gravity \textbf{31}, 165004 (2014).

\bibitem{rev1} K. Hinterbichler, Reviews of Modern Physics \textbf{84}, 671
(2012).

\bibitem{rev2} C. de Rham, Living Reviews in Relativity \textbf{17}, 7
(2014).

\bibitem{rev3} A. Schmidt-May and M. von Strauss, J. Phys. A \textbf{49},
183001 (2016).

\bibitem{Chia} C.-I. Chiang, K. Izumi, and P. Chen, JCAP \textbf{12}, 025
(2012).

\bibitem{cosm1} A. De Felice, A. E. Gumrukcuoglu, C. Lin, and S. Mukohyama,
Class. Quant. Grav. \textbf{30}, 184004 (2013).

\bibitem{cosm2} F. K\"{o}nnig, Y. Akrami, L. Amendola, M. Motta, and A. R.
Solomon, Phys. Rev. \textbf{D 90}, 124014 (2014).

\bibitem{cosm2new} Y. Akrami, S. F. Hassan, F. K\"{o}nnig, A. Schmidt-May, and A. R. Solomon, Phys.Lett. {\bf B 748},  37 (2015).

\bibitem{dR3} C. de Rham, A. J. Tolley, and D. H. Wesley, Phys. Rev. \textbf{%
D 87}, 044025 (2013).

\bibitem{N1} Th. M. Nieuwenhuizen, Phys. Rev. \textbf{D 84}, 024038 (2011).

\bibitem{Cai1} Y.-F. Cai, D. A. Easson, C. Gao, and E. N. Saridakis, Phys.
Rev. \textbf{D 87}, 064001 (2013).

\bibitem{Com} D. Comelli, M. Crisostomi, F. Nesti, and L. Pilo, Phys. Rev.
\textbf{D 85}, 024044 (2012).

\bibitem{Bri} Y. Brihaye and Y. Verbin, Phys. Rev. \textbf{D 86}, 024031
(2012).

\bibitem{Li1} P. Li, X.-Z. Li, and P. Xi, Class. Quantum Grav. \textbf{33},
115004 (2016).

\bibitem{Jaf} G. Jafari, M. R. Setare, and H. R. Bakhtiarizadeh,
arXiv:1702.00189 (2017).

\bibitem{Bur1} P. Burikham, S. Ponglertsakul, and L. Tannukij, Phys. Rev. D
\textbf{96}, 124001 (2017).

\bibitem{bh1} K. Koyama, G. Niz and G. Tasinato, Phys. Rev. D \textbf{84},
064033 (2011).

\bibitem{bh2} K. Koyama, G. Niz and G. Tasinato, Phys. Rev. Lett. \textbf{107%
}, 131101 (2011).

\bibitem{bh3} R. G. Cai, Y. P. Hu, Q. Y. Pan and Y. L. Zhang, Phys. Rev. D
\textbf{91}, 024032 (2015).

\bibitem{bh4} S. H. Hendi, B. E. Panah, and S. Panahiyan, Journal of High Energy Physics {\bf 2015}, 157 (2015).

\bibitem{bh5} S. G. Ghosh, L. Tannukij and P. Wongjun, Eur. Phys. J. C
\textbf{76}, 119 (2016).

\bibitem{bh6} S. H. Hendi, R. B. Mann, S. Panahiyan, and B. Eslam Panah, Phys. Rev. {\bf D 95}, 021501(R) (2017).

\bibitem{bh7} S. H. Hendi, B. Eslam Panah, S. Panahiyan, H. Liu, and X. -H. Meng, Phys. Lett. {\bf B 781}, 40 (2018).

\bibitem{Neut0} T. Katsuragawa, S. Nojiri, S. D. Odintsov, and M. Yamazaki, Phys. Rev. {\bf D 93}, 124013 (2016).

\bibitem{Neut} S. H. Hendi, G. H. Bordbar, B. Eslam Panah, and S. Panahiyan, Journal of Cosmology and Astroparticle Physics {\bf 07},  004  (2017).

\bibitem{Bu59} H. A. Buchdahl, Phys. Rev. \textbf{116}, 1027 (1959).

\bibitem{Stra} N. Straumann, General Relativity and Relativistic
Astrophysics, Springer Verlag, Berlin (1984).

\bibitem{MaDoHa00} M. K. Mak, Peter N. Dobson, Jr., and T. Harko, Mod. Phys.
Lett. A \textbf{15}, 2153 (2000).

\bibitem{An1} H. Andreasson, J. Diff. Eq. \textbf{245}, 2243 (2008).

\bibitem{An2} H. Andreasson, Commun. Math. Phys. \textbf{288}, 715 (2009).

\bibitem{An3} H. Andreasson and C. G. Boehmer, Class. Quantum Grav. \textbf{%
26}, 195007 (2009).

\bibitem{MaDoHa01} M. K. Mak, Peter N. Dobson, Jr., and T. Harko, Europhys.
Lett. \textbf{55}, 310 (2001).

\bibitem{Boehmer:2007gq}  C.~G.~Boehmer and T.~Harko,
Gen.\ Rel.\ Grav.\ \textbf{39}, 757 (2007).

\bibitem{An4} H. Andreasson, C. G. Boehmer, and A. Mussa, Class. Quantum
Grav. \textbf{29}, 095012 (2012).

\bibitem{BoHa06} C. G. Boehmer and T. Harko, Class. Quant. Grav. \textbf{23}%
, 6479 (2006).

\bibitem{W1} M. Wright, Class. Quantum Grav. \textbf{32}, 215005 (2015).

\bibitem{W2} M. Wright, General Relativity and Gravitation \textbf{48}, 1
(2016).

\bibitem{min1} C. G. Boehmer and T. Harko, Phys. Lett. B \textbf{630}, 73
(2005).

\bibitem{min2} C. G. Boehmer and T. Harko, Found. Phys. \textbf{38}, 216
(2008).

\bibitem{min3} C. G. Boehmer and T. Harko, Gen. Rel. Grav. \textbf{39}, 757
(2007).

\bibitem{Wesson:2003qn}  P.~S.~Wesson,  
Mod.\ Phys.\ Lett.\ A \textbf{19}, 1995 (2004).

\bibitem{Not} L. Nottale, Mach’s Principle, Dirac's Large Number Hypothesis and the Cosmological Constant Problem
(preprint), https://www.luth.obspm.fr/ luthier/nottale/arlambda.pdf (1993).

\bibitem{Beck:2008rd}  C.~Beck,
Physica A \textbf{388}, 3384 (2009).

\bibitem{L1} M. J. Lake, J. Phys. Conf. Ser. \textbf{883}, 012001 (2017),
arXiv:1707.07563 [gr-qc]

\bibitem{Mass1} P. Burikham, K. Cheamsawat, T. Harko, and M. J. Lake, Eur.
Phys. J. C \textbf{75}, 442 (2015).

\bibitem{Mass2} P. Burikham, K. Cheamsawat, T. Harko, and M. J. Lake, Eur.
Phys. J. C \textbf{76}, 1-22 (2016).

\bibitem{Mass3} P. Burikham, R. Dhanawittayapol, and T. Wuthicharn,
International Journal of Modern Physics {\bf A 31}, 1650089 (2016).

\bibitem{Mass4} P. Burikham, T. Harko, and M. J. Lake, Phys. Rev. D \textbf{%
94}, 064070 (2016).

\bibitem{Mass5} P. Burikham, T. Harko, and M. J. Lake, The European Physical Journal {\bf C 77}, 803 (2017).

\bibitem{Mass6} C. G. Boehmer, P. Burikham, T. Harko, and M. J. Lake,
The European Physical Journal {\bf C 78},  253 (2018).

\bibitem{Cap1} A. V. Astashenok, S. Capozziello, and S. D. Odintsov, Journal of Cosmology and Astroparticle Physics {\bf 12}, 040 (2013).

\bibitem{Cap2} A. V. Astashenok, S. Capozziello, and S. D. Odintsov, Phys. Rev. {\bf D 89}, 103509 (2014).

\bibitem{Cap3} A. V. Astashenok, S. Capozziello, and S. D. Odintsov, Journal of Cosmology and Astroparticle Physics {\bf 01}, 001 (2015).

\bibitem{Cap4} S. Capozziello, M. De Laurentis, R. Farinelli, and S. D. Odintsov, Phys. Rev. {\bf D 93}, 023501 (2016).

\bibitem{Berezhiani:2011mt}  L.~Berezhiani, G.~Chkareuli, C.~de Rham,
G.~Gabadadze and A.~J.~Tolley,  
Phys.\ Rev.\ D \textbf{85}, 044024 (2012).

\bibitem{Ghosh:2015cva}  S.~G.~Ghosh, L.~Tannukij and P.~Wongjun,
Eur.\ Phys.\ J.\ C \textbf{76}, 119 (2016).

\bibitem{Hasnew}
  S.~F.~Hassan and R.~A.~Rosen,
  JHEP {\bf 1107} (2011) 009
  doi:10.1007/JHEP07(2011)009
  [arXiv:1103.6055 [hep-th]].
  
\bibitem{cosm} A. E. Gumrukcuoglu, C. Lin and S. Mukohyama, JCAP \textbf{1111%
}, 030 (2011).

\bibitem{cosm1a} A. E. Gumrukcuoglu, S. Kuroyanagi, C. Lin, S. Mukohyama and
N. Tanahashi, Class. Quant. Grav. \textbf{29}, 235026 (2012).

\bibitem{cosm2a} C. Lin and M. Sasaki, Phys. Lett. B \textbf{752}, 84 (2016).

\bibitem{cosm3} S. Kuroyanagi, C. Lin, M. Sasaki and S. Tsujikawa,
arXiv:1710.06789 (2017).

\bibitem{cosm4} G. Domenech, T. Hiramatsu, C. Lin, M. Sasaki, M. Shiraishi
and Y. Wang, JCAP \textbf{1705}, 034 (2017).
\\
\end{thebibliography}
\end{document}